\documentstyle[preprint,aps,epsfig]{revtex}

\draft

\preprint{\vbox{\hsize=120pt\noindent SLAC--PUB--8618 \\
October 2000
}}

\begin{document}
%--------------------------------------- Above Title ------------------
%--------------------------------------- Main Title -------------------
\title{
An Improved Direct Measurement of Leptonic Coupling Asymmetries with
Polarized $Z$ Bosons$^\dagger$
}
%--------------------------------------- Author Name ------------------
\author{The SLD Collaboration$^*$}
\address{Stanford Linear Accelerator Center\\
         Stanford University, Stanford, California, 94309\\}
%\date{\today}
\maketitle
%--------------------------------------- Abstract ---------------------
\begin{abstract}
We present final measurements of the $Z$ boson-lepton coupling 
asymmetry parameters $A_e$, $A_{\mu}$, and $A_{\tau}$
with the complete sample of polarized $Z$ bosons collected by the SLD detector 
at the SLAC Linear Collider.
From the left-right production and decay polar angle asymmetries in leptonic
$Z$ decays we measure
$A_e      = 0.1544 \pm 0.0060$, 
$A_{\mu}  = 0.142  \pm 0.015 $, and
$A_{\tau} = 0.136  \pm 0.015 $.
Combined with our left-right asymmetry measured from hadronic decays, we find
$A_e = 0.1516 \pm 0.0021$.
%The $A_e$ result is combined with the left-right asymmetry measurement
%using $Z$ decays to
%hadrons, $A^0_{LR}(\equiv A_e)$, and is found to be $A_e = 0.1516 \pm 0.0021$.
Assuming lepton universality, we obtain a combined effective weak mixing
angle of $\sin^2\theta^{eff}_W = 0.23098 \pm 0.00026$.
\end{abstract}
%
% insert suggested PACS numbers in braces on next line
%\pacs{13.38.Dg,12.15.Ji,13.10.+q}
% *********************************
% * This is for the preprint only *
% *********************************
\vskip 0.3in
\begin{center}
{\rm Submitted to {\em Physical Review Letters}}
\end{center}
\vskip 1.0in
\vbox{\footnotesize\renewcommand{\baselinestretch}{1}\noindent
 $^\dagger$This work was supported in part by Department of Energy
  contract DE-AC03-76SF00515}
\pagebreak
%\bigskip
%--------------------------------------- End of Title Page ------------
%
% author list for inclusion in LaTeX documents
% using \author{} and \address{} commands
%
% Institution number definitions:
%
\begin{center}
\def\iAOMORI{$^{(1)}$}
\def\iBRI{$^{(2)}$}
\def\iBRUN{$^{(3)}$}
\def\iBU{$^{(4)}$}
\def\iCOLO{$^{(5)}$}
\def\iCSU{$^{(6)}$}
\def\iFERR{$^{(7)}$}
\def\iFRAS{$^{(8)}$}
\def\iJHU{$^{(9)}$}
\def\iLBL{$^{(10)}$}
\def\iMASS{$^{(11)}$}
\def\iMISSI{$^{(12)}$}
\def\iMIT{$^{(13)}$}
\def\iMOSCOW{$^{(14)}$}
\def\iNAGO{$^{(15)}$}
\def\iOREG{$^{(16)}$}
\def\iOXF{$^{(17)}$}
\def\iPERU{$^{(18)}$}
\def\iRAL{$^{(19)}$}
\def\iRUTG{$^{(20)}$}
\def\iSLAC{$^{(21)}$}
\def\iSOONG{$^{(22)}$}
\def\iTENN{$^{(23)}$}
\def\iTOHO{$^{(24)}$}
\def\iUCSB{$^{(25)}$}
\def\iUCSC{$^{(26)}$}
\def\iVAND{$^{(27)}$}
\def\iWASH{$^{(28)}$}
\def\iWISC{$^{(29)}$}
\def\iYALE{$^{(30)}$}

\baselineskip=.75\baselineskip 

\mbox{Koya Abe\unskip,\iTOHO}
\mbox{Kenji Abe\unskip,\iNAGO}
\mbox{T. Abe\unskip,\iSLAC}
\mbox{I. Adam\unskip,\iSLAC}
\mbox{H. Akimoto\unskip,\iSLAC}
\mbox{D. Aston\unskip,\iSLAC}
\mbox{K.G. Baird\unskip,\iMASS}
\mbox{C. Baltay\unskip,\iYALE}
\mbox{H.R. Band\unskip,\iWISC}
\mbox{T.L. Barklow\unskip,\iSLAC}
\mbox{J.M. Bauer\unskip,\iMISSI}
\mbox{G. Bellodi\unskip,\iOXF}
\mbox{R. Berger\unskip,\iSLAC}
\mbox{G. Blaylock\unskip,\iMASS}
\mbox{J.R. Bogart\unskip,\iSLAC}
\mbox{G.R. Bower\unskip,\iSLAC}
\mbox{J.E. Brau\unskip,\iOREG}
\mbox{M. Breidenbach\unskip,\iSLAC}
\mbox{W.M. Bugg\unskip,\iTENN}
\mbox{D. Burke\unskip,\iSLAC}
\mbox{T.H. Burnett\unskip,\iWASH}
\mbox{P.N. Burrows\unskip,\iOXF}
\mbox{A. Calcaterra\unskip,\iFRAS}
\mbox{R. Cassell\unskip,\iSLAC}
\mbox{A. Chou\unskip,\iSLAC}
\mbox{H.O. Cohn\unskip,\iTENN}
\mbox{J.A. Coller\unskip,\iBU}
\mbox{M.R. Convery\unskip,\iSLAC}
\mbox{V. Cook\unskip,\iWASH}
\mbox{R.F. Cowan\unskip,\iMIT}
\mbox{G. Crawford\unskip,\iSLAC}
\mbox{C.J.S. Damerell\unskip,\iRAL}
\mbox{M. Daoudi\unskip,\iSLAC}
\mbox{N. de Groot\unskip,\iBRI}
\mbox{R. de Sangro\unskip,\iFRAS}
\mbox{D.N. Dong\unskip,\iSLAC}
\mbox{M. Doser\unskip,\iSLAC}
\mbox{R. Dubois\unskip,\iSLAC}
\mbox{I. Erofeeva\unskip,\iMOSCOW}
\mbox{V. Eschenburg\unskip,\iMISSI}
\mbox{E. Etzion\unskip,\iWISC}
\mbox{S. Fahey\unskip,\iCOLO}
\mbox{D. Falciai\unskip,\iFRAS}
\mbox{J.P. Fernandez\unskip,\iUCSC}
\mbox{K. Flood\unskip,\iMASS}
\mbox{R. Frey\unskip,\iOREG}
\mbox{E.L. Hart\unskip,\iTENN}
\mbox{K. Hasuko\unskip,\iTOHO}
\mbox{S.S. Hertzbach\unskip,\iMASS}
\mbox{M.E. Huffer\unskip,\iSLAC}
\mbox{X. Huynh\unskip,\iSLAC}
\mbox{M. Iwasaki\unskip,\iOREG}
\mbox{D.J. Jackson\unskip,\iRAL}
\mbox{P. Jacques\unskip,\iRUTG}
\mbox{J.A. Jaros\unskip,\iSLAC}
\mbox{Z.Y. Jiang\unskip,\iSLAC}
\mbox{A.S. Johnson\unskip,\iSLAC}
\mbox{J.R. Johnson\unskip,\iWISC}
\mbox{R. Kajikawa\unskip,\iNAGO}
\mbox{M. Kalelkar\unskip,\iRUTG}
\mbox{H.J. Kang\unskip,\iRUTG}
\mbox{R.R. Kofler\unskip,\iMASS}
\mbox{R.S. Kroeger\unskip,\iMISSI}
\mbox{M. Langston\unskip,\iOREG}
\mbox{D.W.G. Leith\unskip,\iSLAC}
\mbox{V. Lia\unskip,\iMIT}
\mbox{C. Lin\unskip,\iMASS}
\mbox{G. Mancinelli\unskip,\iRUTG}
\mbox{S. Manly\unskip,\iYALE}
\mbox{G. Mantovani\unskip,\iPERU}
\mbox{T.W. Markiewicz\unskip,\iSLAC}
\mbox{T. Maruyama\unskip,\iSLAC}
\mbox{A.K. McKemey\unskip,\iBRUN}
\mbox{R. Messner\unskip,\iSLAC}
\mbox{K.C. Moffeit\unskip,\iSLAC}
\mbox{T.B. Moore\unskip,\iYALE}
\mbox{M. Morii\unskip,\iSLAC}
\mbox{D. Muller\unskip,\iSLAC}
\mbox{V. Murzin\unskip,\iMOSCOW}
\mbox{S. Narita\unskip,\iTOHO}
\mbox{U. Nauenberg\unskip,\iCOLO}
\mbox{H. Neal\unskip,\iYALE}
\mbox{G. Nesom\unskip,\iOXF}
\mbox{N. Oishi\unskip,\iNAGO}
\mbox{D. Onoprienko\unskip,\iTENN}
\mbox{L.S. Osborne\unskip,\iMIT}
\mbox{R.S. Panvini\unskip,\iVAND}
\mbox{C.H. Park\unskip,\iSOONG}
\mbox{I. Peruzzi\unskip,\iFRAS}
\mbox{M. Piccolo\unskip,\iFRAS}
\mbox{L. Piemontese\unskip,\iFERR}
\mbox{R.J. Plano\unskip,\iRUTG}
\mbox{R. Prepost\unskip,\iWISC}
\mbox{C.Y. Prescott\unskip,\iSLAC}
\mbox{B.N. Ratcliff\unskip,\iSLAC}
\mbox{J. Reidy\unskip,\iMISSI}
\mbox{P.L. Reinertsen\unskip,\iUCSC}
\mbox{L.S. Rochester\unskip,\iSLAC}
\mbox{P.C. Rowson\unskip,\iSLAC}
\mbox{J.J. Russell\unskip,\iSLAC}
\mbox{O.H. Saxton\unskip,\iSLAC}
\mbox{T. Schalk\unskip,\iUCSC}
\mbox{B.A. Schumm\unskip,\iUCSC}
\mbox{J. Schwiening\unskip,\iSLAC}
\mbox{V.V. Serbo\unskip,\iSLAC}
\mbox{G. Shapiro\unskip,\iLBL}
\mbox{N.B. Sinev\unskip,\iOREG}
\mbox{J.A. Snyder\unskip,\iYALE}
\mbox{H. Staengle\unskip,\iCSU}
\mbox{A. Stahl\unskip,\iSLAC}
\mbox{P. Stamer\unskip,\iRUTG}
\mbox{H. Steiner\unskip,\iLBL}
\mbox{D. Su\unskip,\iSLAC}
\mbox{F. Suekane\unskip,\iTOHO}
\mbox{A. Sugiyama\unskip,\iNAGO}
\mbox{S. Suzuki\unskip,\iNAGO}
\mbox{M. Swartz\unskip,\iJHU}
\mbox{F.E. Taylor\unskip,\iMIT}
\mbox{J. Thom\unskip,\iSLAC}
\mbox{E. Torrence\unskip,\iMIT}
\mbox{T. Usher\unskip,\iSLAC}
\mbox{J. Va'vra\unskip,\iSLAC}
\mbox{R. Verdier\unskip,\iMIT}
\mbox{D.L. Wagner\unskip,\iCOLO}
\mbox{A.P. Waite\unskip,\iSLAC}
\mbox{S. Walston\unskip,\iOREG}
\mbox{A.W. Weidemann\unskip,\iTENN}
\mbox{E.R. Weiss\unskip,\iWASH}
\mbox{J.S. Whitaker\unskip,\iBU}
\mbox{S.H. Williams\unskip,\iSLAC}
\mbox{S. Willocq\unskip,\iMASS}
\mbox{R.J. Wilson\unskip,\iCSU}
\mbox{W.J. Wisniewski\unskip,\iSLAC}
\mbox{J.L. Wittlin\unskip,\iMASS}
\mbox{M. Woods\unskip,\iSLAC}
\mbox{T.R. Wright\unskip,\iWISC}
\mbox{R.K. Yamamoto\unskip,\iMIT}
\mbox{J. Yashima\unskip,\iTOHO}
\mbox{S.J. Yellin\unskip,\iUCSB}
\mbox{C.C. Young\unskip,\iSLAC}
\mbox{H. Yuta\unskip.\iAOMORI}

\it
\vskip \baselineskip                   % \bigskip did not work
%\centerline{(The SLD Collaboration)}   % include collaboration name
\vskip \baselineskip
\baselineskip=.75\baselineskip   % shrink the interline spacing
\iAOMORI
  Aomori University, Aomori , 030 Japan, \break
\iBRI
  University of Bristol, Bristol, United Kingdom, \break
\iBRUN
  Brunel University, Uxbridge, Middlesex, UB8 3PH United Kingdom, \break
\iBU
  Boston University, Boston, Massachusetts 02215, \break
\iCOLO
  University of Colorado, Boulder, Colorado 80309, \break
\iCSU
  Colorado State University, Ft. Collins, Colorado 80523, \break
\iFERR
  INFN Sezione di Ferrara and Universita di Ferrara, I-44100 Ferrara, Italy, \break
\iFRAS
  INFN Lab. Nazionali di Frascati, I-00044 Frascati, Italy, \break
\iJHU
  Johns Hopkins University,  Baltimore, Maryland 21218-2686, \break
\iLBL
  Lawrence Berkeley Laboratory, University of California, Berkeley, California 94720, \break
\iMASS
  University of Massachusetts, Amherst, Massachusetts 01003, \break
\iMISSI
  University of Mississippi, University, Mississippi 38677, \break
\iMIT
  Massachusetts Institute of Technology, Cambridge, Massachusetts 02139, \break
\iMOSCOW
  Institute of Nuclear Physics, Moscow State University, 119899, Moscow Russia, \break
\iNAGO
  Nagoya University, Chikusa-ku, Nagoya, 464 Japan, \break
\iOREG
  University of Oregon, Eugene, Oregon 97403, \break
\iOXF
  Oxford University, Oxford, OX1 3RH, United Kingdom, \break
\iPERU
  INFN Sezione di Perugia and Universita di Perugia, I-06100 Perugia, Italy, \break
\iRAL
  Rutherford Appleton Laboratory, Chilton, Didcot, Oxon OX11 0QX United Kingdom, \break
\iRUTG
  Rutgers University, Piscataway, New Jersey 08855, \break
\iSLAC
  Stanford Linear Accelerator Center, Stanford University, Stanford, California 94309, \break
\iSOONG
  Soongsil University, Seoul, Korea 156-743, \break
\iTENN
  University of Tennessee, Knoxville, Tennessee 37996, \break
\iTOHO
  Tohoku University, Sendai 980, Japan, \break
\iUCSB
  University of California at Santa Barbara, Santa Barbara, California 93106, \break
\iUCSC
  University of California at Santa Cruz, Santa Cruz, California 95064, \break
\iVAND
  Vanderbilt University, Nashville,Tennessee 37235, \break
\iWASH
  University of Washington, Seattle, Washington 98105, \break
\iWISC
  University of Wisconsin, Madison,Wisconsin 53706, \break
\iYALE
  Yale University, New Haven, Connecticut 06511. \break

\rm
%
%  }   % end of address list

\end{center}

%============================ TOF LOF LOT =============================
%%% comment/uncomment to your liking
%\tableofcontents
%\newpage
%\bigskip
%\listoffigures
%\newpage
%\listoftables
%
%============================ Main Text ===============================
%%%%%%%%%%%%%%%
%% Introduction
%%%%%%%%%%%%%%%

The extent of parity violation in the electroweak interaction can
be probed directly in the production and decay of polarized $Z$ bosons
generated by $e^+e^-$ annihilation.
Parity violation in $Z$ production ($e^+e^- \to Z$) and decay into 
charge lepton pairs ($Z \to e^+e^-, \mu^+\mu^-, \tau^+\tau^-$)
is characterized by the $Z$ boson-lepton coupling asymmetry parameters
$A_e$, $A_{\mu}$, and $A_{\tau}$. 
The asymmetry parameter is defined as 
$A_l = 2v_l a_l/(v_l^2+a_l^2)$,
where $v_l$ and $a_l$ are the effective vector and axial-vector couplings
of the $Z$ boson to the lepton (flavor ``$l$'') current,
respectively.
The Standard Model (SM) assumes lepton universality, 
so that all three species of leptonic asymmetry parameters
are expected to be identical and directly related to 
the effective electroweak mixing angle ($\sin^2\theta^{eff}_W$),
$A_l = 2(1-4\sin^2\theta^{eff}_W)/[1+(1-4\sin^2\theta^{eff}_W)^2]$.
The effective electroweak mixing angle depends on virtual electroweak 
radiative corrections including those which involve the Higgs boson 
and those arising from new phenomena outside of the scope of 
the SM.
Presently, the most stringent upper bounds on the SM Higgs mass are provided
by measurements of $\sin^2\theta^{eff}_W$.

The SLAC Linear Collider (SLC) produces polarized $Z$ bosons
in $e^+e^-$ collisions at the $Z$ resonance
using a longitudinally polarized electron beam.
Electron polarization ($P_e$) allows us to form 
the left-right cross-section asymmetry
to extract the initial state asymmetry parameter $A_e$~\cite{Abe:2000dq}
and also enables us to directly measure the final state asymmetry parameter
$A_l$ for lepton $l$
using the left-right forward-backward asymmetry~\cite{Abe:1997xm}
($\tilde{A}_{FB}=\frac{3}{4}|P_e|A_l$).
Experiments at the $Z$ resonance
without beam polarization~\cite{:2000nr}
have measured the product of initial and
final state asymmetry parameters ($A_{FB}=\frac{3}{4}A_e\cdot A_l$).
Those same experiments have also measured the tau 
polarization~\cite{:2000nr}
which yields $A_e$ and $A_{\tau}$ separately.
The SLC beam polarization enables us to
present the only direct measurement of $A_\mu$.
With 75\% beam polarization, 
the left-right forward-backward asymmetries yield a statistical precision 
equivalent to measurements using a 25 times larger event sample
with the unpolarized forward-backward asymmetry. 

In this letter,
we report new results on direct measurements of the asymmetry parameters
$A_e$, $A_{\mu}$, and $A_{\tau}$ using leptonic $Z$ decays.
The measurements are based on the $3.8 \times 10^5Z$s
collected during 1996-98
by the SLAC Large Detector (SLD) experiment at the SLC.
These results are combined with 
earlier leptonic asymmetry measurements~\cite{Abe:1997xm} 
and the more precise left-right asymmetry measurement using $Z$ decays to 
hadrons~\cite{Abe:2000dq},
to give final measurements
based on the complete sample of polarized $Z$ bosons.

%%%%%%%%%%%%%%%
%% SLC and SLD
%%%%%%%%%%%%%%%

This analysis relies on the Compton 
polarimeter~\cite{Abe:2000dq,polarimeter}, 
tracking by the vertex detector and 
the central drift chamber (CDC)~\cite{Fero:1995pv},
and the liquid argon calorimeter (LAC)~\cite{Axen:1993dz}.
Details about the SLC, the polarized electron source, 
and SLC operation with a polarized beam
can be found in Ref.~\cite{Woods:1996ph}.
Only the details most relevant to this analysis are mentioned here.

In our previous measurements~\cite{Abe:1997xm}, 
the analysis was restricted to the polar-angle range of
$|\cos\theta|<0.7$ due to decreasing tracking and trigger efficiency
for muon-pair final states beyond this region,
even though the high $|\cos\theta|$ region is very 
sensitive to the asymmetry parameters.
In 1996 we installed an upgraded vertex detector (VXD3)~\cite{Abe:1997bu}
and a new trigger system for forward muon pair events.
The improved acceptance of VXD3 allows highly efficient track finding up to 
$|\cos\theta|=0.9$~\cite{Abe:2000ky}.
The new trigger for $\mu^+\mu^-$ events
covers the angular range up to $|\cos\theta|<0.95$
by requiring two back-to-back tracks that pass through the interaction point 
and reach the endcap Warm Iron Calorimeter~\cite{Benvenuti:1989kh}.

%%%%%%%%%%%%%%%
%% THEORY
%%%%%%%%%%%%%%%

Polarization-dependent lepton asymmetries are easily computed from 
$e_{L,R}^- + e^+ \to Z^0 \to l^- + l^+$,
where $l$ represents an electron, a muon, or a tau lepton.
The differential cross section is expressed as follows~\footnote{
For $P_e$, we use the convention that left-handed bunches have negative sign.
}:
%
% Equation
%
\[
\frac{d}{dx}\sigma_{Z}(x,s,P_e;A_e,A_l)
\equiv f_{Z}(s)\Omega_{Z}(x,P_e;A_e,A_l)
= f_{Z}(s)\left[ (1-P_eA_e)(1+x^2)+(A_e-P_e)A_l 2x \right],
%\label{Eq:zterm}
\]
where $s$ is the squared center-of-mass energy and 
$x=\cos\theta$ gives the direction of the outgoing lepton ($l^-$) 
with respect to the electron-beam direction.
Photon exchange terms and, if the final state leptons are electrons,
$t$-channel contributions have to be taken into account.
The leptonic asymmetry parameters which refer to the initial and final state
lepton appear in this expression as $A_e$ and $A_l$, respectively. 
It was determined that $|P_e|=76.16 \pm 0.40$\% and $72.92 \pm 0.38$\%
for the 1996 and 1997-98 runs, respectively~\cite{Abe:2000dq}.

%%%%%%%%%%%%%%%
%% EVENT SELECTION
%%%%%%%%%%%%%%%

%Leptonic $Z$ decays are identified by the characteristic 
%low charged multiplicity and two back-to-back leptons.
%Lepton-pair events are required to have between 2 and 8 
%charged tracks, each of which must pass
%within 1 cm of the nominal $e^+e^-$ interaction point. 
Leptonic $Z$ decay candidates are required to have between 2 and 8 
charged tracks, each of which must pass
within 1 cm of the nominal $e^+e^-$ interaction point.
This excludes most hadronic $Z$ decays, which have an average
charged multiplicity of approximately 20. 
%Since the leptons
%have about 45 GeV in energy, there is little problem assigning
%reconstructed tracks to one of two event hemispheres,
%corresponding to the two leptons. 
One hemisphere must have a net
charge 1 and the other a net charge -1 to ensure unambiguous
assignment of the scattering
angle. 
Each event is assigned a polar-production angle 
with respect to the electron beam direction
based on the thrust axis ($\cos\theta_{thrust}$) 
defined by the charged tracks and we require 
$|\cos\theta_{thrust}| < 0.9$ (0.8) for 1997-98 (96) data.
%Additional
%requirements are imposed to select $e^+e^-$, $\mu^+\mu^-$, and
%$\tau^+\tau^-$ final states and further reduce backgrounds.

%
%Electron event selection
%
A single additional cut is required to select the $e^+e^-$ final state.
We consider the highest-momentum track in each hemisphere and require
the sum of the associated energies deposited in the LAC to exceed 45 GeV.
The $e^+e^-$ candidates have a small contamination (0.7\%)
from $\tau^+\tau^-$ events.

%
%Muon event selection
%
For events of the type $Z\rightarrow\mu^+\mu^-$, we require
the invariant mass of the charged tracks (assumed pion mass) be greater than
70 GeV/c$^2$. 
This removes most $Z\rightarrow\tau^+\tau^-$ events
and virtually all two-photon and hadronic $Z$ decay events.
%The majority of the remaining events are $Z\rightarrow\mu^+\mu^-$ and
%$e^+e^-\rightarrow e^+e^-$. 
We remove the $e^+e^-$ final state by
requiring the energy deposited in the LAC by the highest momentum
track in each hemisphere to be less than 10 GeV.
The muon-pair sample has a very small contamination (0.2\%) from
$\tau^+\tau^-$ final states.

%
%Tau event selection
%
The tau-pair final state selection requires
the event mass to be less than 70 GeV/c$^2$ to remove $\mu^+\mu^-$ final 
states.
The maximum energy per hemisphere in the LAC associated
to a charged track is required to be less than 27 GeV (23 GeV) for
$\cos\theta<0.7$ ($>0.7$) to reject $e^+e^-$ final states.
Two-photon events are suppressed by requiring the angle between 
the total track momenta of the
two hemispheres be greater than $160^\circ$
and by requiring one charged
track to have momentum greater than 4 GeV/c. 
The remaining background from hadronic $Z$ decays is suppressed
by requiring each hemisphere invariant mass, measured using charged
tracks, to be less than 1.6 GeV/c$^2$. The tau-pair candidates
have some contamination from muon pair (2.9\%), electron pair (0.9\%), 
two-photon events (0.9\%), and hadronic final states (0.6\%).

\begin{figure}[p]
\centering
\epsfysize=5.0in
\centerline{\epsffile{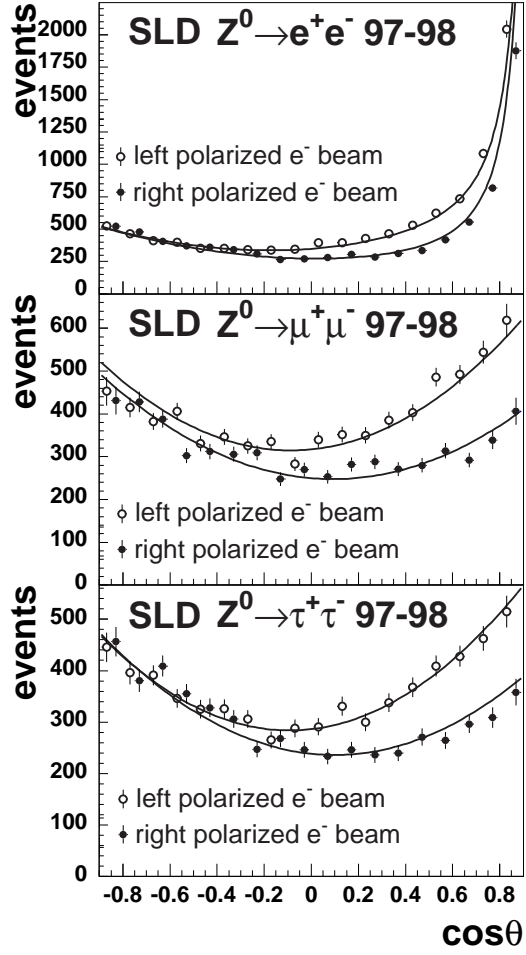}}
\caption{
Polar-angle distributions for $Z$ decays to $e$, $\mu$ and $\tau$
pairs for the 1997-98 SLD run. 
The solid line represents the fit, 
while the points with error bars show the data in bins of 0.1
in $\cos\theta_{thrust}$.
For $|\cos\theta_{thrust}|>0.7$, 
the data are corrected for a decrease in
the detection efficiency with increasing $|\cos\theta_{thrust}|$.}
\label{Fig:distr}
\end{figure}

%
%Selected candidate events
%
Table~\ref{Table:selection} summarizes the selection efficiencies,
backgrounds and numbers of selected candidates
for $e^+e^-$, $\mu^+\mu^-$, and $\tau^+\tau^-$ final states.
Fig.~\ref{Fig:distr} shows the $\cos\theta$ distributions for $e^+e^-$,
$\mu^+\mu^-$, and $\tau^+\tau^-$ candidates for the 1997-98 data.  
The asymmetries in the 1996 data are similar
but have smaller acceptance ($|\cos\theta| \le 0.8$).
%The solid line represents the fit, 
%while the points with error bars show the data in bins of 0.1
%in $\cos\theta$.

%%%%%%%%%%%%%%%
%% Maximum Likelihood Method
%%%%%%%%%%%%%%%
%Simple asymmetries can be used to extract
%$A_e$ and $A_l$ from the data. 
%The left-right asymmetry measures
%the difference in $Z$ production for the left- and right-handed polarized
%electron beams. 
%The left-right forward-backward asymmetry is a
%double asymmetry which is formed by taking the difference in
%the number of forward ($F$) and backward ($B$) events for left
%($L$) and right ($R$) beam polarization. 
%Here forward (backward) means $\cos\theta > 0$ ($\cos\theta < 0$).
%These asymmetries can be derived from Eq.~(\ref{Eq:zterm}), 
%using obvious subscripts, 
%(assuming uniform acceptance over the full solid angle 
% in the case of $\tilde{A}^l_{FB}$):
%%
%% Equation
%%
%\begin{eqnarray}
%A_{LR}&=&\frac{1}{|P_e|}\frac{N_L-N_R}{N_L+N_R}=A_e \label{Eq:alr}
%\\
%\tilde{A}^l_{FB}&=&\frac{4}{3}\frac{1}{|P_e|}
%	\frac{(N_{LF}-N_{LB})-(N_{RF}-N_{RB})}
%	{(N_{LF}+N_{LB})+(N_{RF}+N_{RB})}=A_l.
%\label{Eq:apfb}
%\end{eqnarray}
%%

%The essence of the measurement is expressed in
%Eqs.~(\ref{Eq:alr}) and (\ref{Eq:apfb}), but instead of simply
We perform a maximum likelihood fit, event by event, 
to incorporate the contributions of all the terms in the cross section and to
include the effect of initial state radiation.
%The maximum likelihood fit is less sensitive to detector
%acceptance as a function of polar angle than the counting method,
%and has more statistical power.
We define 3 likelihood functions for individual lepton final states.
$A_e$ and $A_{\mu}$ ($A_{\tau}$) are derived from $\mu^+\mu^-$ 
($\tau^+\tau^-$) final states.
These $A_e$ results are combined with the number obtained from $e^+e^-$ 
final states.

The likelihood function for muon- and tau-pair final states is defined as follows:
%
% Equation
%
\begin{equation}
{\cal L}(x,s,P_e;A_e,A_l)=
\int ds^\prime 
H(s,s^\prime)
\left\{
	\frac{d}{dx}\sigma_{Z}(x,s^\prime,P_e;A_e,A_l) 
	+ \frac{d}{dx}\sigma_{Z\gamma}(x,s^\prime,P_e;A_e,A_l)
	+ \frac{d}{dx}\sigma_{\gamma}(x,s^\prime)
\right\} ,\label{Eq:likelihood function(MIZA1)}
\end{equation}
where $A_e$ and $A_l$(=$A_{\mu}$ or $A_{\tau}$) are free parameters
and $H(s,s^\prime)$ is a radiator function.
The integration over $s^\prime$ is done with the program
MIZA~\cite{Martinez:1991ta} to take into account the initial state
radiation.
The spread in the beam energy has a negligible effect.
$(d\sigma_{Z}/dx)(...)$, $(d\sigma_{\gamma}/dx)(...)$, 
and $(d\sigma_{Z\gamma}/dx)(...)$ are the tree-level differential
cross sections for $Z$ exchange, photon exchange, and their interference.
The integration is performed before the fit
to obtain the coefficients $\bar{f}_Z$, $\bar{f}_{Z\gamma}$, 
and $\bar{f}_\gamma$, and the likelihood function becomes
%
% Equation
%
\begin{equation}
{\cal L}(x,s,P_e;A_e,A_l)=
 	\bar{f}_Z(s) \Omega_{Z}(x,P_e;A_e,A_l)
	+ \bar{f}_{Z\gamma}(s) \Omega_{Z\gamma}(x,P_e;A_e,A_l)
	+ \bar{f}_{\gamma} (s) \Omega_{\gamma}(x). 
\label{Eq:likelihood function(MIZA2)}
\end{equation}
These coefficients give the relative sizes of the three terms at
the SLC center-of-mass energy 
($\sqrt{s}=91.237 \pm 0.029$ GeV for the 1997-98 run 
and $91.26 \pm 0.03$ GeV for 1996)~\cite{Abe:2000dq}.

The $e^+e^-$ final state includes
both $s$-channel and $t$-channel $Z$ and photon exchanges
which yields four amplitudes and ten cross-section terms.
All ten terms are energy-dependent.
We define a maximum likelihood function for $e^+e^-$ final states
by modifying Eqs.~(\ref{Eq:likelihood function(MIZA1)}) and 
(\ref{Eq:likelihood function(MIZA2)})
to include all ten terms.
The integration over $s^{\prime}$ is performed with 
DMIBA~\cite{Martinez:1992wy} to obtain the coefficients
for the relative size of the ten terms. 

%%%%%%%%%%%%%%%
%% Systematic Effects
%%%%%%%%%%%%%%%
There are several systematic effects which can bias the results. 
The uncertainties associated with these effects are summarized in
Table~\ref{Table:systematics} and are small compared with the statistical
uncertainties.
The uncertainty on the beam polarization is correlated among all the
measurements and corresponds to an uncertainty on 
$A_l$ of $\pm 0.0008$.
The uncertainty in the amount of background and its effect on the
fitted parameters are taken into account.
The background contaminations have been derived from
detailed Monte Carlo simulations as well as from studying the effect of
cuts in background-rich samples of real data. 
The uncertainty in the asymmetry parameters 
due to a $\pm1\sigma$ variation of $\sqrt{s}$ 
(which affects radiative corrections)
is of the order $10^{-4}$, except for the $A_e$ determination from $e^+e^-$ 
final states for which it is of order $10^{-3}$.

%
% V-A
%
The dominant systematic error in the tau analysis results from 
the V-A structure of tau decay~\cite{Tsai:1971vv}, which introduces 
a selection bias in our analysis.
For example, if both taus decay to $\pi\nu$, 
helicity conservation requires that both pions generally have 
lower momentum for a left-handed $\tau^-$ and right-handed $\tau^+$ and
higher momentum otherwise.
This effect, which biases the reconstructed event mass, is large at
the SLD because the high beam polarization induces a very high and
asymmetric tau polarization as a function of polar angle.
Using detailed Monte Carlo simulation~\cite{Jadach:1994yv,Jadach:1993hs}, 
we find an overall shift in $A_{\tau}$
of $+0.0182\pm0.0018$ ($+0.0183 \pm 0.0017$) for the 1997-98 (1996) runs
due to the effect of the V-A structure,
where the uncertainty is from Monte Carlo statistics.
The value extracted from the fit must be reduced by this amount.
The value of $A_e$ extracted from $\tau^+\tau^-$ final states is
not affected since the overall relative efficiencies for
left-handed beam and right-handed beam events are not changed significantly
(only the polar angle dependence of the efficiencies are
changed). 

Tracks are less well measured at very high $|\cos\theta|$
and charge confusion for these tracks
dilutes the asymmetries.
We estimate this effect by comparing the numbers of opposite sign
back-to-back tracks with same-sign pairs.
The uncertainty is found to be $\pm 0.0007$ and $\pm 0.0011$ for $A_{\mu}$ 
and $A_{\tau}$, respectively.
A small detector-induced forward-backward asymmetry
would also introduce a small bias for $A_{\tau}$.
Using a two-photon enriched data sample,
we find a small forward-backward asymmetry effect in the momentum distribution
of negatively-signed charged tracks ($\sim 1.0$ GeV/c).
We estimate this causes a systematic uncertainty of $\pm0.0004$ for 
$A_{\tau}$,
while the effect is negligible for $A_e$ and $A_{\mu}$.
The selection efficiency as a function of polar
angle is another possible source of bias in $A_l$.
If this efficiency is symmetric about $\cos\theta=0$ then
$A_l$ is unaffected for muons and taus.
%(see Eqs.~(\ref{Eq:alr}) and (\ref{Eq:apfb})). 
However, the maximum
likelihood fit for the $e^+e^-$ final state may be affected even for
a symmetric efficiency, if it is not uniform. 
This systematic uncertainty is estimated to be 
$\pm0.0002$ for $A_e$ by using the Monte Carlo simulation to compare 
the nominal result with the result for 100\% selection efficiency for 
the $e^+e^-$ final state.
We have also studied the effect of the uncertainty in the
thrust axis determination, which also includes the uncertainty from 
the final state radiation,
and found that the contribution is negligible. 

%%%%%%%%%%%%%%%
%% Results for 96-98 SLD runs
%%%%%%%%%%%%%%%

We find the results for $A_e$, $A_{\mu}$, and $A_{\tau}$
using the 1996-98 SLD runs to be
$A_e      = 0.1549 \pm 0.0066 (stat.) \pm 0.0013 (syst.)$,
$A_{\mu}  = 0.152  \pm 0.016  (stat.) \pm 0.001  (syst.)$, and
$A_{\tau} = 0.121  \pm 0.017  (stat.) \pm 0.003  (syst.)$, respectively.
%
%%%%%%%%%%%%%%%
%% Combined SLD results on Z-Lepton couplings
%%%%%%%%%%%%%%%
%
We combine these results with
our previous leptonic asymmetry measurements~\cite{Abe:1997xm},
accounting for small effects due to correlations in systematic uncertainties 
(polarization and average SLD center-of-mass energy).
From purely leptonic final states, we obtain
$A_e=0.1544 \pm 0.0060$.
We also combine the $A_e$ result with the left-right asymmetry measurement
using $Z$ decays to hadrons ($A_{LR}^0\equiv A_e$)~\cite{Abe:2000dq}
and obtain:
%
% Equation
%
$$
\begin{array}{cclcll}
A_{e}    & = & 0.1516 &\pm& 0.0021 \  (\mbox{with}\ A_{LR}^0)& ;\cr
A_{\mu}  & = & 0.142  &\pm& 0.015 & ; \mbox{and}\cr
A_{\tau} & = & 0.136  &\pm& 0.015 .&
\end{array}
$$
Our results are consistent with lepton universality. 
Assuming universality, we combine these results into $A_{l}$,
which in the context of the standard model is simply related to the
electroweak mixing angle:
%
% Equation
%
$$
A_{l} = 0.15130 \pm 0.00207 \qquad 
	\sin^2\theta_{W}^{eff}=0.23098\pm0.00026 .
$$
%
% SM Higgs...
%
Within the context of the SM, the result above can be used to constrain the
mass of the Higgs boson.
We use the measured $Z$ boson~\cite{:2000nr} and 
top quark~\cite{Cite:Top} masses, a determination of 
$\alpha(M_Z^2)$~\cite{Martin:2000by},
and the ZFITTER~6.23 program~\cite{Bardin:1999yd}
to obtain a 95\% confidence level upper bound of 147 GeV/c$^2$.

%%%%%%%%%%%%%%%
%% Conclusion
%%%%%%%%%%%%%%%

In conclusion,
we have presented direct measurements of the $Z$ boson-lepton coupling
asymmetries $A_e$, $A_\mu$, and $A_\tau$ using
$e^+e^-\rightarrow e^+e^-$, $\mu^+\mu^-$, $\tau^+\tau^-$ events
produced with a longitudinally polarized electron beam
during the 1996-98 SLD runs.
These results are combined with our
previously
published results, yielding SLD's final result for the weak mixing
angle.  
This is presently the most precise available determination of this quantity.

%%%%%%%%%%%%%%%
%% Acknowledge
%%%%%%%%%%%%%%%

We thank the personnel of the SLAC accelerator department and the
technical staffs of our collaborating institutions for their outstanding
efforts on our behalf. This work was supported by the Department of
Energy, the National Science Foundation, the Istituto Nazionale di
Fisica Nucleare of Italy, the Japan-US Cooperative Research Project
on High Energy Physics, and the Science and Engineering Research
Council of the United Kingdom.

% now the references. delete or change fake bibitem. delete next three
%   lines and directly read in your .bbl file if you use bibtex.

% tables follow here
%
% Here is an example of the general form of a table:
% Fill in the caption in the braces of the \caption{} command. Put the label
% that you will use with \ref{} command in the braces of the \label{} command.
% Insert the column specifiers (l, r, c, d, etc.) in the empty braces of the
% \begin{tabular}{} command.
%
% \begin{table}
% \caption{}
% \label{}
% \begin{tabular}{}
% \end{tabular}
% \end{table}

\begin{table}[hbtp]
\caption{Summary of event selections, efficiency, and purity for 
$Z\rightarrow l^+l^-$ for the 1997-98 (1996) data.
%Results for 1996 data are given in parentheses.
}
%e^+e^-$,
%$Z\rightarrow\mu^+\mu^-$ and $Z\rightarrow\tau^+\tau^-$.}
\label{Table:selection}
\begin{center}
\begin{tabular}{llcl}
Event  & Background as \%  & Efficiency in            & \# of Selected \\
Sample & of Selected Events & $|\cos\theta|<0.9$ ($|\cos\theta|<0.8$)
	& Events\\ \hline
$e^+e^-\rightarrow e^+e^-$ &
   0.7\% (0.8\%)$\tau^+\tau^-$  & 75\% (87\%) & 15675(2052) \\ \hline
$Z\rightarrow\mu^+\mu^-$   &
   0.2\% (0.2\%) $\tau^+\tau^-$ & 77\% (83\%) & 11431(1625) \\ \hline
$Z\rightarrow\tau^+\tau^-$ & 
   0.9\% (0.7\%) $e^+e^-$        &             & \cr
 & 2.9\% (2.2\%) $\mu^+\mu^-$    & 70\% (77\%)& 10841(1494) \\
 & 0.9\% (0.9\%) two-photon      &             & \\
 & 0.6\% (0.3\%) hadrons         &             & \\
\end{tabular}
\end{center}
\end{table}

\begin{table}[h]
\caption{\label{Table:Systematic-Error} 
Summary of statistical and systematic uncertainties in units of $10^{-4}$ 
for the 1997-98 (1996) data.
}
\label{Table:systematics}
\begin{center}
\begin{tabular}{lccccc}
Source & $A_e^e$ & $A_e^{\mu}$ & $A_e^{\tau}$ & $A_{\mu}^{\mu}$ &
$A_{\tau}^{\tau}$   \\ \hline
Statistics & 110(280) & 130(330) & 130(340) & 180(470) & 180(480) \\
Polarization            &  8 (8) &  8 (8)&  8 (8) &  8 (8) &  8 (8)\\
Backgrounds             &  5 (3) & --    & 13 (14)& --     & 14 (13)\\
Radiative Correction    & 23 (17)&  2 (2)&  2  (2)&  3 (1) &  2 (2)\\
V-A                     & --     & --    & --     & --     & 18 (17)\\
Charge Confusion        & --     & --    & --     &  7 (--)& 11 (1) \\
Detector asymmetry      & --     & --    & --     & --     & 4 (4)\\
Nonuniform efficiency   & 2 (--) & --    & --     & --     & -- 
\end{tabular}
\end{center}
\end{table}

% figures follow here
%
% Here is an example of the general form of a figure:
% Fill in the caption in the braces of the \caption{} command. Put the label
% that you will use with \ref{} command in the braces of the \label{} command.
%
% \begin{figure}
% \caption{}
% \label{}
% \end{figure}

\clearpage

\end{document}